\documentclass[a4paper,11pt]{article}
\pdfoutput=1 % if your are submitting a pdflatex (i.e. if you have
\usepackage{jheppub} % for details on the use of the package, please
\usepackage[T1]{fontenc} % if needed
\usepackage{mathtools}
\usepackage{romannum}
\usepackage[normalem]{ulem}

\title{\boldmath Reheating in Models with Non-minimal Coupling \\
in metric and Palatini formalisms}

\author[a]{Dhong Yeon Cheong,}
\author[a]{Sung Mook Lee,}%,\note{DYC, SML: co-first authors, SCP: corresponding author}}
\author[a, b]{and Seong Chan Park} 

\affiliation[a]{Department of Physics and IPAP, Yonsei University, Seoul 03722, Republic of Korea}
\affiliation[b]{Korea Institute for Advanced Study, Seoul 02455, Republic of Korea}

\emailAdd{dhongyeon@yonsei.ac.kr}
\emailAdd{sungmook.lee@yonsei.ac.kr}
\emailAdd{sc.park@yonsei.ac.kr}

\abstract{We study reheating of inflationary models with general non-minimal coupling $K(\phi)R$ with $K(\phi)\sim \sqrt{V(\phi)}$ where $R$ is the Ricci scalar and $V$ is the inflaton potential. In particular, when we take the monomial potential $K(\phi) \propto \phi^m$ with $m \in \mathbb{Z}_+$, we provide general analytic expressions for cosmological observables. We consider a wide range of non-minimal coupling $ \xi \in  [0, \infty) $ in metric and Palatini formalisms and derive the predictions for cosmological observables and the reheating temperature taking a general equation of state parameter $ w_\text{reh} $.}

\begin{document} 
\maketitle
\flushbottom

\section{Introduction}
Cosmological inflation~\cite{Guth:1980zm, Linde:1981mu,Albrecht:1982wi}, now widely accepted as a `standard procedure' in early universe, has been heavily tested with increasing precision~\cite{WMAP:2010qai, Planck:2018jri, BICEPKeck:2021gln}. However, the particle physics details governing reheating after inflation, which ultimately provides the initial conditions for the hot big bang, is largely unknown. Reheating involves the physics in the perturbative and non-perturbative decay, resonances and nonlinear dynamics of inflaton~\cite{Dolgov:1989us, Traschen:1990sw, Kofman:1994rk, Shtanov:1994ce, Kofman:1997yn}.

Phenomenologically, the reheating stage is effectively parameterized by the phenomenological parameters $(T_{\text{reh}}, N_{\text{reh}}, w_{\text{reh}}) $ where $T_{\rm reh}$ is the reheating temperature,  $ N_{\text{reh}} $ is the e-folding number, and $w_\text{reh}$ is the (constant) effective equation of state during reheating, respectively. In general, direct cosmological observables are hardly traceable up to the period of reheating, but some indirect bounds are available especially in the single field inflation models: taking the CMB data into account, consistency relations among the parameters $(T_{\text{reh}}, N_{\text{reh}}, w_{\text{reh}}) $ can be derived~\cite{Cook:2015vqa, Drewes:2015coa}. Numerous studies have been conducted along this line~\cite{Cai:2015soa, Rehagen:2015zma, Ellis:2015pla, deFreitas:2015xxa, Ueno:2016dim, Dalianis:2016wpu, Drewes:2017fmn, Krajewski:2018moi, Rashidi:2018ois, DiMarco:2018bnw, Sakhi:2019vcx, Gialamas:2019nly, Asadi:2019iod, German:2020iwg, Das:2020kff, Haro:2020qos, Kanfon:2020yxx}.

Among the various models compatible with observations that one may consider, models with a non-minimal coupling between the Ricci scalar $ R $ and the inflaton field $ \phi $ are highly motivated \cite{Futamase:1987ua, Fakir:1990eg, Komatsu:1999mt, Park:2008hz, Hertzberg:2010dc, Kim:2013ehu, Park:2018kst}. It is known that a generic class of cases exist where the non-minimal coupling ($K(\phi) R$) induces the asymptotic flatness ($\frac{V(\phi)}{K(\phi)^2} \sim \text{const.}$), hence guarantees approximate shift symmetry of the inflationary potential and suppresses the tensor-to-scalar ratio $ r $~\cite{Park:2008hz, Park:2008tj}. This class is also known as the $\alpha$-attractor~\cite{Kallosh:2013yoa}. and it is being constrained by precision CMB measurements (i.e. Planck, WMAP, COBE and BICEP/Keck), and the detection of $ r $ is one of the main targets of several future experiments including CMB-S4 and LiteBIRD \cite{Abazajian:2019eic,LiteBIRD:2020khw}. One notable example in the inflationary model class with a non-minimal coupling is the Higgs inflation, where the Standard Model (SM) Higgs $ h $ has the role of the inflaton with a non-minimal coupling $ K(h) = \xi h^{2}$ and a quartic potential $ V(h) = \frac{\lambda}{4} h^{4} $ \cite{Bezrukov:2007ep}, especially in the vicinity of a critical point ~\cite{Hamada:2014wna, Hamada:2014iga}. (See Ref.~\cite{Cheong:2021vdb} for a recent review.)

Although various theoretical and phenomenological issues of non-minimally coupled inflation models have been explored \cite{Futamase:1987ua,Fakir:1990eg,Makino:1991sg,Komatsu:1999mt,Park:2008hz,Hertzberg:2010dc,Linde:2011nh,Jinno:2019und,Lee:2020yaj}, in this work, we further refine the cosmological predictions of these models implementing effects of reheating. Reheating is often implicitly assumed to be nearly instantaneous, and conventional wisdom for $\alpha$-attractor class of models (including $R^2$-driven Starobinsky model) says the cosmological observables are degenerate among different realizations, but its specific reheating dynamics may depend on microscopic details and alter the observational predictions \cite{Chung:1998rq,Giudice:2000ex,Davidson:2000er,Harigaya:2013vwa}. Considering various theoretical possibilities, we want to cover  directions worth pursuing as follows:
\begin{itemize} 
	\item {\bf Metric and Palatini formalism:} In the Palatini formalism of gravity, the affine connection $ \Gamma^{\rho}_{\mu\nu} $ and the metric $ g_{\mu\nu} $ are introduced in an independent manner. Although the Palatini formalism is equivalent to the metric formalism at the level of pure Einstein-Hilbert action, it is known that they provide different predictions when non-minimal couplings are present \cite{Bauer:2008zj, Tamanini:2010uq, Borowiec:2011wd,  Tenkanen:2017jih, Racioppi:2017spw, Jarv:2017azx, Racioppi:2018zoy, Jinno:2019und, Tenkanen:2020dge, Jarv:2020qqm}.

	\item {\bf General  monomial  potentials:}  A generic monomial potential $V(\phi) \propto K(\phi)^2 \propto \phi^{2m} $  guarantees the asymptotic flat potential in the Einstein frame~\cite{Park:2008hz}:
	\begin{align} \label{Eq:generalcondition}
		\lim_{\phi \rightarrow \infty} \left(\frac{V}{K^{2}}\right) = \text{Const.}>0.
	\end{align}
	The condition essentially describes the $\alpha$-attractor behavior~\cite{Kallosh:2013yoa}. 
	This criterion is applied in the metric formalism as well as Palatini cases.

	\item {\bf  Wide range of non-minimal coupling $ \xi \in [0, \infty) $:} Even though large non-minimal couplings are usually considered, a wide range of $ \xi $ (including ones much smaller than unity) consistent to observations still remains valid. We consider general cases including $ \xi = 0 $ (minimal, monomial) and $ \xi = \infty $ (maximally non-minimal) \cite{Cook:2015vqa}.
\end{itemize}

This paper is organized as follows. In Section.~\ref{Section:Inflationary and Reheating Predictions}, we set our model in both metric and Palatini formalism of gravity and provide the inflationary predictions. In Section.~\ref{Section:Consistency Relations between Reheating Predictions}, we relate the cosmological observables $ (n_{s},r) $ and reheating parameters. We also discuss current/future bound on reheating temperatures from these observables. We conclude in Section.~\ref{Section:Conclusion and Discussions}.

\section{Model} \label{Section:Inflationary and Reheating Predictions}

In this section, we set the inflationary model with non-minimal coupling term in metric and Palatini formalism.

\subsection{Non-minimal Coupling in metric and Palatini formalisms}

As the Einstein-Hilbert action already includes the $(Mass)^2$ dimensional coupling to the Ricci scalar $R$, we are enforced to include the operators of the form $\hat{\cal O} R$ with $[\hat{\cal O}] = (Mass)^2$ following the effective field theory approach unless a symmetry forbids it.  Therefore when we introduce a scalar field $\phi$ to our theory, we include an arbitrary (gauge invariant) function $K(\phi)$ directly coupling to gravity:
\begin{align}
S_J = \int d^{4}x \sqrt{-g_{J}} \left[ - \frac{M^{2} + K(\phi)}{2} R +  \frac{1}{2} (\partial \phi)^{2} - V(\phi)	 \right],
\end{align}
with $[K]=(Mass)^2$ and an arbitrary mass parameter $ M $. The potential $V(\phi)$ is a function of $[V]=(Mass)^4$. In metric formalism, the Ricci scalar $ R $ is solely determined by the metric $g_{\mu\nu}$, but in the Palatini formalism it is also determined by the connection $\Gamma^\mu_{\nu\lambda}$ which is taken to be independent of the metric. In the present of non-minimal coupling $K(\phi)R$, the two formalisms are not equivalent and predicts different Universe \cite{Bauer:2008zj}. 

By performing the Weyl transformation,
\begin{align}
g_{E,\mu\nu} = \Omega^{2} g_{J,\mu\nu},&&\Omega^{2} \equiv \frac{M^{2} + K(\phi)}{M_{P}^{2}}
\end{align}
the action is conveniently transformed into the canonical form of gravity in Einstein frame
\begin{align}
S = \int d^{4}x \sqrt{-g_{E}} \left[ - \frac{M_{P}^{2}}{2}R_{E} + \frac{1}{2} \Pi(\phi) (\partial \phi)^{2} - \frac{V(\phi)}{\Omega^{4}} \right],
\end{align}
where the non-trivial kinetic term is given by 
\begin{align}
\Pi(\phi)  \equiv \frac{1}{ \Omega^{2}}  +  \frac{3\zeta}{2 M_{P}^{2}} \frac{K^{\prime} (\phi)^{2}}{\Omega^{4}},
&& \zeta = \begin{dcases}
1 & \text{(Metric) } \\
0 & \text{(Palatini) } 
\end{dcases}.
 \label{eq:2eq}
\end{align}
in metric and Palatini formalism, respectively. The second term ($\propto K'^2$) originates from the transformation of the Ricci scalar $ R(\Gamma) $, making it  absent in the Palatini formalism.
The kinetic term is easily canonicalized by $ \sqrt{\Pi(\phi)} \partial h = \partial \phi$ or 
\begin{align} \label{Eq:JordantoEinstein}
\frac{d h}{d \phi} = \sqrt{\Pi(\phi)}.
\end{align}
When the explicit function $K(\phi)$ is given, we obtain the canonical scalar $h(\phi)$ by integration.

An asymptotically flat, positive potential in the Einstein frame $V_E (h(\phi))$ should satisfy a general condition at large fields~\cite{Park:2008hz}:
\begin{align}
\lim_{\phi \to \infty} \frac{V(\phi)}{K(\phi)^2} = \text{Const.} >0. 
\end{align}
We request this condition $V \sim K^2$ or equivalently $ K \sim \sqrt{V}$ to realize successful slow-roll inflation. We emphasize again that the condition essentially describes the $\alpha$-attractor behavior~\cite{Kallosh:2013yoa}.

\subsection{Monomial functions: $K(\phi) \sim \phi^m \sim \sqrt{V(\phi)}$}

To explicitly see the inflationary behavior, we choose $ K $ and $ V $ to be monomial functions with $m\geq 1$~\footnote{For models with a global $U(1)$ symmetry, a $K(\phi)$ function with an odd power of $\phi$ is usually forbidden.} 
\begin{align}
K(\phi ) = \xi M_{P}^{2}\left(\frac{\phi}{M_{P}}\right)^{m},&&	V = \frac{\lambda M_{P}^{4}}{2m} \left(\frac{\phi}{M_{P}}\right)^{2m},
\label{Eq:KandV}
\end{align}
where $ \xi $ and $ \lambda $ are arbitrary dimensionless parameters. In addition, without a (or with a sufficiently small) vacuum expectation value of the field $ \phi $, we have $ M \simeq M_{P} $ to guarantee the canonicalized Einstein frame field (later we {define} as $ h $) and potential $ U(\phi(h)) $ coincide with the Jordan frame field $ \phi $ and potential $ V(\phi) $ near the origin. Therefore, we set $ M = M_{P} $ without losing precision in our predictions. Then the potential in Einstein frame becomes
\begin{align} \label{Eq:Potential}
U(\phi) 
= \frac{M_{P}^{4}}{\left(M_{P}^{2} + K\right)^{2}} V 
= \frac{\lambda M_{P}^{4} \left(\frac{\phi}{M_{P}}\right)^{2m}}{2m\left( 1 + \xi \left(\frac{\phi}{M_{P}}\right)^{m}  \right)^{2}}.
\end{align}
The slow-roll parameters are defined in the Einstein frame:
\begin{align}
\epsilon(\phi) 
&\equiv \frac{M_P^2}{2} \left(\frac{\partial U(\phi(h))/\partial h}{U}\right)^2 = \frac{M_P^2}{2\Pi_\zeta(\phi)} \left(\frac{\partial U/\partial \phi}{U}\right)^2, \\
\eta(\phi) &\equiv M_P^2 \frac{\partial^2 U/\partial h^2}{U} = \frac{M_P^2}{\sqrt{\Pi_\zeta(\phi)} U} \frac{\partial}{\partial \phi} \left(\frac{1}{\sqrt{\Pi_\zeta(\phi)}} \frac{\partial U}{\partial \phi}\right), 
\end{align}
where $\Pi_\zeta$ is defined in Eq.~\eqref{eq:2eq} for metric ($\zeta=1$) and Palatini ($\zeta=0$) cases with $ K(\phi) $ in Eq.~\eqref{Eq:KandV} and $x=\phi/M_P$:
\begin{align}
\Pi_\zeta(\phi = M_P x) = \frac{1+\xi x^m + \zeta \frac{3 m^2\xi^2}{2}x^{2(m-1)}}{(1+\xi x^m)^2}, && \zeta = \begin{dcases}
1 & \text{(Metric) } \\
0 & \text{(Palatini) } 
\end{dcases}.
\end{align} 
The explicit form of the slow-roll parameters are obtained 
\begin{align} \label{Eq:exactepsilon}
\epsilon(\phi=M_P x) &= \frac{4 m^2 }{2 x^2 \left(\xi x^m+1\right)+3 \zeta  m^2 \xi ^2 x^{2 m}},  \\ \label{Eq:exacteta}
\eta(\phi=M_P x) & = \frac{4 m \left[x^2 \left(\xi  x^m+1\right) \left\{ (m+2) \xi  x^m-4
   m+2\right\} -3 \zeta  m^3 \xi ^2 x^{2 m}
   \left(\xi  x^m-1\right)\right]}{\left[2 x^2 \left(\xi x^m+1\right)+3 \zeta  m^2 \xi ^2 x^{2 m}\right]^2}.
\end{align}
It is noticed that the slow-roll parameters are independent of $\lambda$. 

The cosmological time (in Einstein frame) $ t_{k} $ when the mode corresponding to the pivot scale $ k $ leaves the horizon is determined by $ k = a(t_{k}) H(t_{k}) $ with the scale factor $ a $ and $ H $ being the Hubble parameter. In this work, the pivot scale is chosen to be $k = 0.05 \text{Mpc}^{-1}$. On the other hand, the time $ t_{e} $ at the end of the inflation is set by $\epsilon(\phi(t_{e})) =1$.

Then, the number of e-foldings during expansion from a pivot scale with $ a_{k} = a(t_{k})$ to the end of inflation at $ a_{e} = a(t_{e}) $ is given by
\begin{align}
\label{Eq:exactefold}
N(\phi_k) \equiv N_k
&= \log \frac{a_e}{a_k} =\frac{1}{M_P^2} \int_{h(\phi_e)}^{h(\phi_k)} dh  \frac{ U  }{\partial U/\partial h} \nonumber  \\
&= \int_{\phi_e}^{\phi_k}  \frac{d\phi}{M_P} \sqrt{\frac{\Pi_\zeta(\phi)}{2\epsilon(\phi)}} 
 = \left[\frac{x^2}{4m} + \frac{3}{4}\zeta \left(\xi x^m -  \ln(M_P^m + \zeta M_P^m x^m )\right)\right]_{\phi_e/M_P}^{\phi_k/M_P} 
\end{align}
with $ \phi_k \equiv \phi(t_k) $, $ \phi_e \equiv \phi(t_e)$, and $ H_{k} \equiv H(t_{k}) $ where we used $dh = d\phi \sqrt{\Pi_\zeta}$ in the 2nd line.

\subsection{$ \xi \ll 1 $ (Metric $\approx$ Palatini) }
We first analyze the $\xi\ll1$ limit close to the minimal case. The exact slow-roll parameters Eq.~(\ref{Eq:exactepsilon}) and  Eq.~(\ref{Eq:exacteta}) in this limit then take the approximate expression 
\begin{align}
\epsilon(\phi = M_P x) & \simeq \frac{2m^2}{x^2} - 2  m^2 \xi x^{m-2} + \left(2 x^2 - 3 m^2 \zeta \right)  m^2 \xi^2 x^{2m-4} + \mathcal{O}(\xi^3), \\
\eta(\phi=M_P x) & \simeq \frac{2m (2m-1)}{x^2} - 5  m^2 \xi x^{m-2} + \left[5x^2 +3(2-3m)m\zeta\right] m^2 \xi^2 x^{2m-4}+\mathcal{O}(\xi^3)
\end{align}
where $\zeta = 0,1$ each correspond to the Palatini and metric cases, respectively. Note that the $\zeta$ parameter dependence appears in the $\mathcal{O}(\xi^2)$ order, i.e. the predictions of both cases will deviate in $\mathcal{O}(\xi^2)$, and the observables approximately coincide for $\xi \ll 1$.

The field value where inflation ends for both cases is determined by $\epsilon =1$, hence 
\begin{align}
\phi_e \simeq \sqrt{2} m M_P - 2^{\frac{m-1}{2}}m^{1+m} M_P \xi  + 2^{m-\frac{5}{2}} m^{2 m+1} ( 2 m+3-3 \zeta) M_P \xi ^2  +\mathcal{O}(\xi^3). 
\end{align}
Consequently, the e-folding number $N_k$ from the pivot scale $a_k$ to the end of inflation $ a_{e} $ then becomes, following Eq.~(\ref{Eq:exactefold})
\begin{align}
N_k \simeq \frac{\phi_k^2 - \phi_e^2 }{4 m M_P^2} + \frac{3}{8} \zeta \left[ \frac{\phi_k^{2m} -\phi_e^{2m} }{M_P^{2m}}\right] \xi^2. 
\end{align}  
As $\xi \ll 1$, the field value $\phi_k$ at pivot scale is represented in terms of $N_k$ as
\begin{align}
\phi_k \simeq \sqrt{2 m (m+2 N_k)}M_P  -\frac{2^{\frac{m}{2}-\frac{1}{2}} m^{m+2} }{\sqrt{m (m+2
   N_k)}} M_P  \xi + \mathcal{O}(\xi^2)
\end{align}
leading to the slow-roll parameters
\begin{align}
	\epsilon\left(\phi_k\right) &\simeq \frac{m}{m+2 N_k} +\frac{2^{\frac{m}{2}} \left[m^{m+2}-m^{\frac{m}{2}+1} (m+2N_k)^{\frac{m}{2}+1}\right] \xi}{(m+2 N_k)^2}  + \mathcal{O}(\xi^2)  ,  \\ 
	\eta\left(\phi_k \right) & \simeq \frac{2m-1}{2 N_k+m}+\frac{2^{\frac{m}{2}-1} \left[2 m^{m+1} (2m-1)-5 m^{\frac{m}{2}+1} (m+2N_k)^{\frac{m}{2}+1}\right]  \xi}{(m+2 N_k)^2}+ \mathcal{O}( \xi^2). 
\end{align}
These expressions yield the spectral index $n_s$ and the tensor-to-scalar ratio $r$ in this limit as
\begin{align}
	n_s & \simeq  1 - 6 \epsilon + 2 \eta \vert_{\phi = \phi_{k}} \\
	& \simeq 1-\frac{2 (m+1)}{m+2 N_k} +\frac{2^{\frac{m}{2}} \left[m^{\frac{m}{2}+1} (m+2 N_k)^{\frac{m}{2}+1}-2 m^{m+1} (m+1)\right]\xi}{(m+2 N_k)^2} + \mathcal{O}(\xi^2)   \\ 
	r & \simeq 16 \epsilon \vert_{\phi = \phi_{k}} \simeq \frac{16 m}{m+2 N_k} +\frac{2^{\frac{m}{2} + 4}  \left[m^{m+2}-m^{\frac{m}{2}+1} (m+2N_k)^{\frac{m}{2}+1}\right] \xi}{(m+2 N_k)^2}  + \mathcal{O}(\xi^2) 
\end{align}
along with the curvature power spectrum amplitude being 
\begin{align}
A_s &\simeq \left.\frac{1}{ 24 \pi^2 \epsilon} \frac{U}{M_P^4}\right|_{\phi = \phi_k} 
\\ & \simeq 
\frac{ 2^{m-4} m^{m-2} (m+2 N_k)^{m+1} \lambda  }{3 \pi ^2 } \nonumber \\
& \qquad - \frac{2^{\frac{3 m}{2}-4} m^{\frac{3m}{2}-2} \lambda   \left[ (m+1)m^{\frac{m}{2}+1} (m+2 N_k)^m+(m+2 N_k)^{\frac{3m}{2}+1}\right]  \xi}{3 \pi ^2 }
\end{align}
where $ A_{s} \simeq 2.1 \times 10^{-9} $ is given by Planck and BICEP/Keck 2018 results at the pivot scale $k = 0.05 \text{Mpc}^{-1}$ \cite{Planck:2018jri}. The $\zeta$ dependent terms are suppressed, with its leading order terms emerging at $\mathcal{O} (\xi^2)$ order.

Note that, the specific $\xi$ range that resembles an approximate minimal coupling case differs for each $m$ value. In order for a noticeable deviation of $ (n_{s},r) $ from minimal cases, the required $\delta \xi $ involves 
\begin{align}
 \left\vert \frac{\delta r}{r} \right\vert \sim \left\vert \frac{\delta n_{s}}{n_{s}} \right\vert  \sim (4 m {N_k})^{\frac{m}{2}} \vert \delta \xi \vert 
\end{align}
which is $m$ dependent. A larger power monomial potential will lead to a smaller $\delta \xi$ that shows a reasonable deviation from $\xi=0$ predictions.

In the minimal coupling case $(\xi = 0 )$, the relations of the inflationary observables simplify to 
\begin{align}
	n_{s} \simeq 1 - \frac{m+1}{N_{k}},&&
	r \simeq \frac{8m}{N_{k}}, 
\end{align}
and these two parameters are correlated as 
\begin{align}
	N_{k} \simeq \frac{m+1}{1-n_{s}} \simeq \frac{8m}{r} && \Leftrightarrow  && r \simeq \frac{8m}{m+1} (1-n_s) \approx \left.\frac{0.32 m}{m+1}\right|_{n_s\approx 0.96}. 
\end{align}
The resulting value of $r$ is a bit too large  compared to the observed limit $r \lesssim 0.1$~\cite{WMAP:2010qai, Planck:2018jri, BICEPKeck:2021gln}.

\subsection{$ \xi \gg 1$ (Metric $\neq$ Palatini)}
We now turn to the opposite limit and examine the case with a large non-vanishing non-minimal coupling. 
Now the predictions from metric and Palatini formalisms drastically differ. Therefore we discuss the metric case and Palatini case below separately. 

\subsubsection{Metric Formalism}

In the large $ \xi $ limit, slow roll parameters are approximated as
\begin{align}
\epsilon &\simeq \frac{4}{3\xi^{2}} \left(\frac{M_{P}}{\phi}\right)^{2m} + \mathcal{O}(\xi^{-3}), \\
\eta &\simeq - \frac{4}{3\xi} \left(\frac{M_{P}}{\phi}\right)^{m} + \frac{4}{3\xi^{2}} \left[\left( \frac{M_{P}}{\phi}  \right)^{2m} + \frac{3m-2}{3m^{3}} \left(\frac{M_{P}}{\phi}\right)^{2m-2}			\right] + \mathcal{O}(\xi^{-3}). 
\end{align}
Therefore, $ \epsilon(\phi_{e}) = 1$ gives
\begin{align} \label{Eq:MetricEndField}
\phi_{e} \simeq \left(\frac{4}{3\xi^{2}}\right)^{\frac{1}{2m}}M_{P}.
\end{align}
E-folding number and the field value at the pivot scale are obtained
\begin{align}
	N_{k} \approx \frac{3\xi}{4} \left(\frac{\phi_{k}}{M_{P}}\right)^{m} && \Rightarrow  &&\phi_{k} \simeq \left( \frac{4N_{k}}{3 \xi}\right)^{\frac{1}{m}}M_{P}, && (\xi \gg 1, ~\text{Metric}).
\end{align}
Then, the slow roll parameters and cosmological observables are represented by $ N_{k} $ as
\begin{align}
\epsilon(\phi_{k}) \simeq \frac{3}{4N_{k}^{2}}, &&
\eta(\phi_{k}) \simeq - \frac{1}{N_{k}} + \frac{3}{4N_{k}^{2}} + \frac{3m-2}{4N_{k}^{2} m^{3}}  \left( \frac{4N_{k}}{3\xi} \right)^{\frac{2}{m}} 
\end{align}
and the spectral index and tensor-to-scalar-ratio are
\begin{align} \label{Eq:SpectralIndex0}
	n_{s} \simeq 1 - 6 \epsilon + 2 \eta \vert_{\phi = \phi_{k}} \simeq
	1 - \frac{2}{N_{k}} - \frac{3}{N_{k}^{2}}, &&
	r_{s} \simeq 16 \epsilon \vert_{\phi = \phi_{k}} \simeq 
	\frac{12}{N_{k}^{2}},
\end{align}
which enjoys the sweet spot of the Planck, WMAP and BICEP/Keck bounds on $ (n_{s},r) $ for $ N_{k} \simeq 50-60 $. Prospects from future measurements are discussed in Section.~\ref{Section:Consistency Relations between Reheating Predictions}.

From the fact that the potential during inflation approaches a constant value at the large field regime with a large non-minimal coupling,
\begin{align}
U(\phi) \xrightarrow[]{\xi \gg 1, ~ \phi \gg M_{P} / \xi^{1/m}} U_{\text{inf}} \simeq \frac{\lambda M_{P}^{4}}{2m \xi^{2}}
\end{align}
we have the normalization of $ \lambda/\xi^{2} $ as
\begin{align}
A_{s} \simeq \left. \frac{1}{24\pi^{2} \epsilon} \frac{U}{M_{P}^{4}} \right\vert_{\phi = \phi_{k} \gg M_{P} / \xi^{1/m}}  \simeq  \frac{N_{k}^{2} \lambda}{36 m \pi^{2} \xi^{2}} \simeq 2.1 \times 10^{-9}.
\end{align}

\subsubsection{Palatini Formalism} 

The slow-roll parameters in Palatini formalism are approximated as
\begin{align}
\epsilon \simeq \dfrac{2 m^{2} M_{P}^{m+2}}{\xi \phi^{m + 2}}, &&
\eta \simeq - \frac{m(m + 2) M_{P}^{2}}{\phi^{2}}.
\end{align}
By requiring $ \epsilon(\phi_{e}) = 1$,
\begin{align} \label{Eq:PalatiniEndField}
\phi_{e} = 2^{\frac{1}{2 + m}} M_{P} \left(\frac{\xi}{m^{2}}\right)^{-\frac{1}{2+m}}.
\end{align}
The e-folding number and the field value at pivot scale are 
\begin{align}
	N_{k}  \simeq \frac{1}{4 m M^{2}} \left( \phi_{k}^{2} - \phi_{e}^{2} \right) 
	&& \Rightarrow && 
	\phi_{k} \simeq 2 \sqrt{m N_{k}} M_{P}, && (\xi \gg 1,~ \text{Palatini})
\end{align}
where we neglect $ \phi_{e} \ll \phi_{k} $. Now slow roll parameters and cosmological observables are represented as
\begin{align}
	\epsilon(\phi_{k})
	 \simeq \frac{m^{1-\frac{m}{2}}}{2^{1+m} N_{k}^{1+\frac{m}{2}}\xi}, &&
	\eta(\phi_{k})
	\simeq - \frac{2+m}{4N_{k}} + \frac{5 m^{1-\frac{m}{2}}}{2^{2+m} N_{k}^{1+\frac{m}{2}} \xi} 
\end{align}
and the spectral index and the tensor-to-scalar ratio are 
\begin{align} 
	n_{s} \simeq
	1 - \frac{2 + m}{2 N_{k}} -  \frac{m^{1-\frac{m}{2}}}{2^{1+m} N_k^{1+\frac{m}{2}}\xi}, &&	      r\simeq 
	\frac{m^{1-\frac{m}{2}}}{2^{m-3} N_{k}^{1+\frac{m}{2}}\xi}.
\end{align}
One should note that $r \propto 1 / \xi $ is highly suppressed by $ \xi \gg 1 $.
In Palatini case, CMB power spectrum normalization determines $ \lambda / \xi $ as
\begin{align}
A_{s} \simeq \left. \frac{1}{24\pi^{2} \epsilon} \frac{U}{M_{P}^{4}} \right\vert_{\phi \simeq \phi_{k}}  
\simeq \frac{  2^{m-3}
	m^{\frac{m}{2}-2}
	N_{k}^{\frac{m}{2}+1}\lambda}{
	3 \pi ^2 \xi }.
\label{Eq:PalaHiggsAmp}
\end{align}
The fact that $ \lambda/\xi $ is normalized requires a larger order of magnitude of $ \xi $ for the same $ \lambda $ compared to metric cases.

\section{Reheating and Cosmological Predictions} \label{Section:Consistency Relations between Reheating Predictions}
\subsection{Reheating}

In this section we enlist the calculations of $N_\text{reh}$ and $T_\text{reh}$ by modeling the reheating epoch to be described with a constant equation of state  $w_\text{reh}$.
Note that the equation of state $w(t) \equiv p / \rho $, in general, is a time dependent parameter determined by the particle physics details of the reheating process~\cite{Podolsky:2005bw, Lozanov:2016hid, Saha:2020bis, DiMarco:2021xzk}. In our approach following Ref.~\cite{Cook:2015vqa},  we rather take $ w_\text{reh} $ as an average value over the reheating
\begin{align}
	w_\text{reh} \equiv  \frac{1}{N_{\text{reh}}} \int_{N_{k}}^{N_{k}+N_{\text{reh}}} w(N) ~dN.
\end{align}
In this work, we allows a wide range of $w_\text{reh} \in [0, 1/3]$. Recall that $w = 0 $ for matter domination and $w = 1/3$ for radiation domination.\footnote{In principle, one might take $ w_\text{reh} \in [-1,1] $ depending on the detail of the model.}

Now we take the standard  assumptions for the late time cosmology after the reheating and get the relations for $T_{\rm reh}$ and $N_{\rm reh}$. First, the e-folds for the epoch is expressed as 
\begin{align}
	N_\text{reh} \equiv \ln \left(\frac{a_\text{reh}}{a_e}\right) = \frac{1}{3(1+w_\text{reh})}\ln \left(\frac{\rho_e}{\rho_\text{reh}}\right),
\end{align}
where $N_\text{reh}$ represents the duration of reheating from the end of inflation, $a_\text{reh}$ and $a_e$ is the scale factor at the end of reheating and the end of inflation, respectively, while $\rho_\text{reh}$ and $\rho_e$ is the corresponding energy density at the end of reheating and the end of inflation. From $\epsilon_H = -\frac{\dot{H}}{H^2} = 1$ at the end of inflation, we have
\begin{align}
	\rho_e = \frac{3}{2} U_e,
\end{align}
hence  $w(t_e) =-1/3$. Also, the radiation energy with the relativistic degrees of freedom $g_\text{reh}$ and the temperature $T_\text{reh}$ at the end of reheating is given by
\begin{align}
	\rho_\text{reh} = \frac{\pi^2}{30} g_\text{reh} T_\text{reh}^4.
\end{align}
Assuming no additional entropy production after reheating, we can link $T_{\rm reh}$ and the temperature of our current universe $T_0$ as
\begin{align}
	T_\text{reh} 
	= T_0 \left(\frac{a_0}{a_\text{reh}}\right)\left(\frac{g_{0}}{g_{\text{reh}}}\right)^{1/3}=\left(\frac{43}{11 g_{\text{reh}}}\right)^{1/3} \left(\frac{a_0 T_0}{k}\right)H_k e^{-N_k}e^{-N_\text{reh}}
\end{align}
where $g_{0}=2 + \frac{7}{8}\times N_\text{eff} \times \frac{4}{11} $ with $N_\text{eff} =3.046$ in the SM is the relativistic degree of freedom at current universe and $g_{\text{reh}}$ is that at the end of reheating, respectively. 
In the second line, we take the length of e-folds of radiation dominance, $N_{\rm RD} = \ln (a_{\rm eq}/a_{\rm reh})$ with the scale factor at the matter-radiation equality $a_{\rm eq}$, and the ratio $a_0/a_{\rm eq} = a_0 (H_k/k) e^{-N_k} e^{-N_{\rm reh}} e^{-N_{\rm RD}}$ from the pivot scale $k=a_k H_k$. 

Combining the expressions, we obtain $N_\text{reh}$ as
\begin{align}
	N_\text{reh} = \frac{4}{3(1+w_\text{reh})} \left[\frac{1}{4}\ln \left(\frac{45}{\pi^2 g_\text{reh}}\right) + \ln \left(\frac{U_e^{1/4}}{H_k} \right) + \frac{1}{3} \ln \left(\frac{11g_\text{reh}}{43}\right)
	\right. \nonumber \qquad\qquad \\ß
	\left.+\ln \left(\frac{k}{a_0 T_0}\right) + N_k + N_\text{reh}\right].
\end{align}
Taking the standard values $ g_\text{reh} = 106.75$, $k = 0.05~\text{Mpc}^{-1}$, $T_0 = 2.725 ~\text{K}$ for $w_\text{reh} \neq -1/3$ we arrive at
\begin{align}
	N_\text{reh} = \frac{4}{(1-3w_\text{reh})} \left[ 61.6 - \ln \left(\frac{U_e^{1/4}}{H_k}\right) - N_k\right],
\end{align}
and accordingly, inserting this expression to $T_\text{reh}$ gives 
\begin{align}
	T_\text{reh} = \left[\left(\frac{43}{11 g_\text{reh}}\right)^{1/3} \frac{a_0 T_0}{k} H_k e^{-N_k} \left(\frac{45 U_e}{\pi^2 g_\text{reh}}\right)^{-\frac{1}{3(1+w_\text{reh})}}\right]^{\frac{3(1+w_\text{reh})}{3w_\text{reh}-1}}.
\end{align}

\subsection{Results : Observables \& Reheating Temperature }
\label{Section:Results_section}

\begin{figure}[t]
	\centering
	\includegraphics[width=7.5cm]{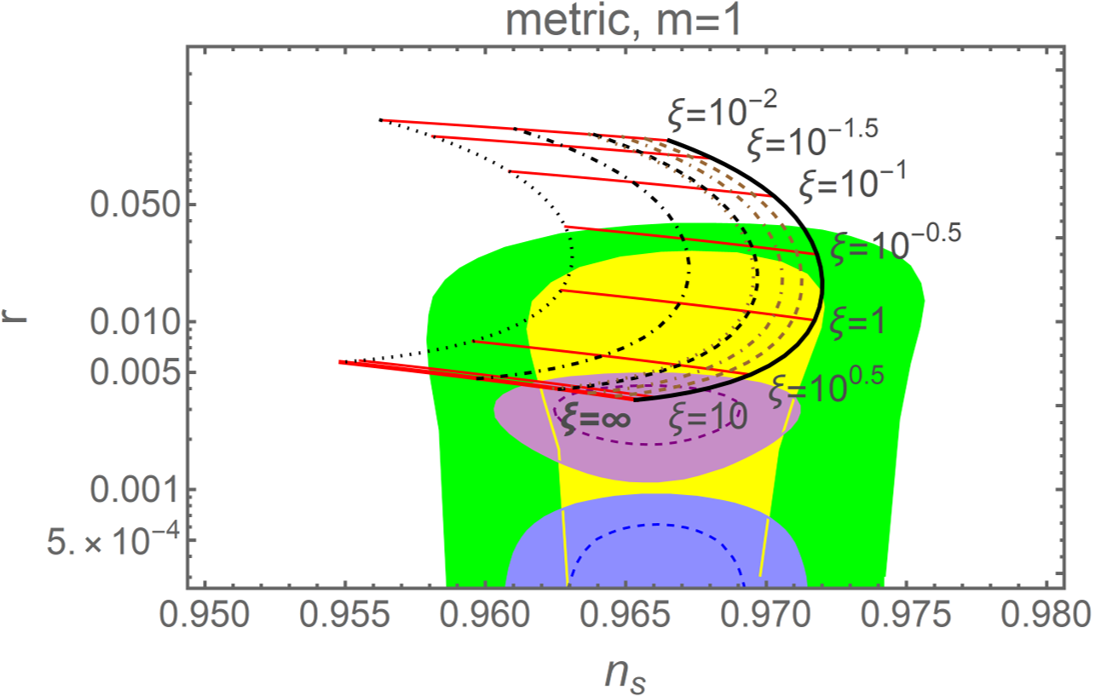}
	\includegraphics[width=7.5cm]{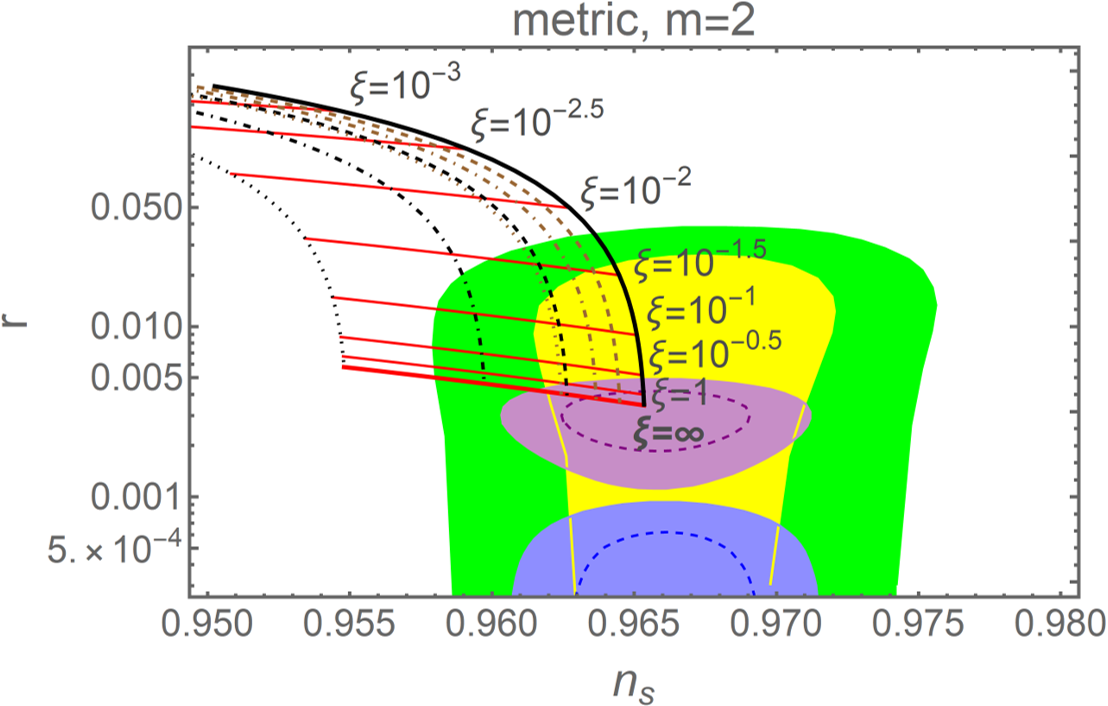}
	\includegraphics[width=7.5cm]{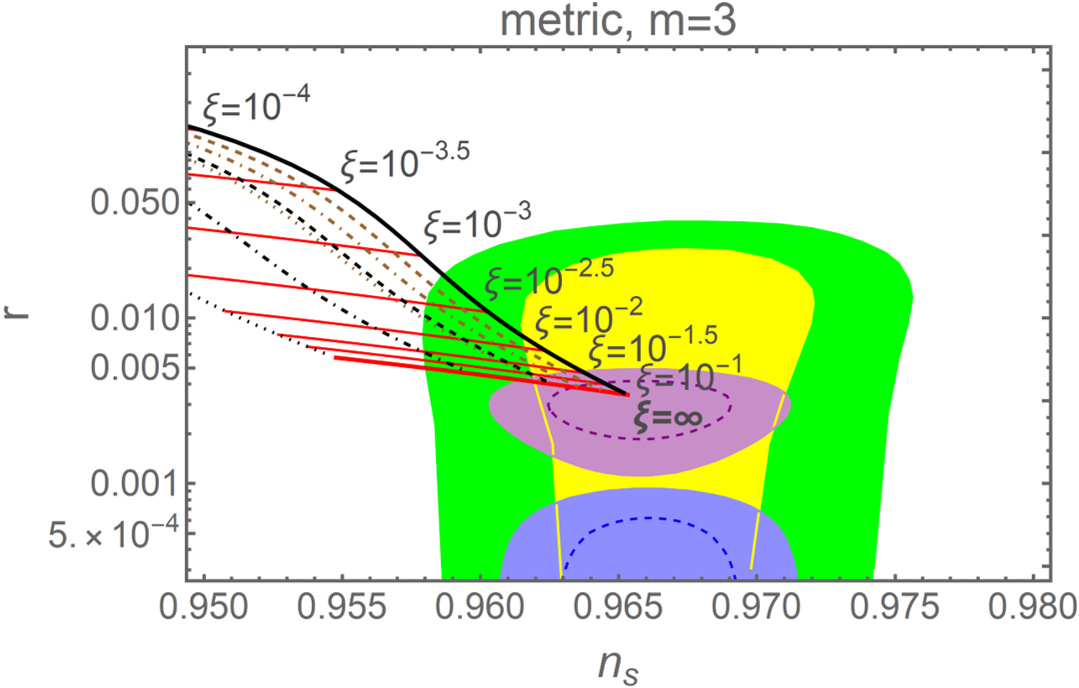}
	\includegraphics[width=7.5cm]{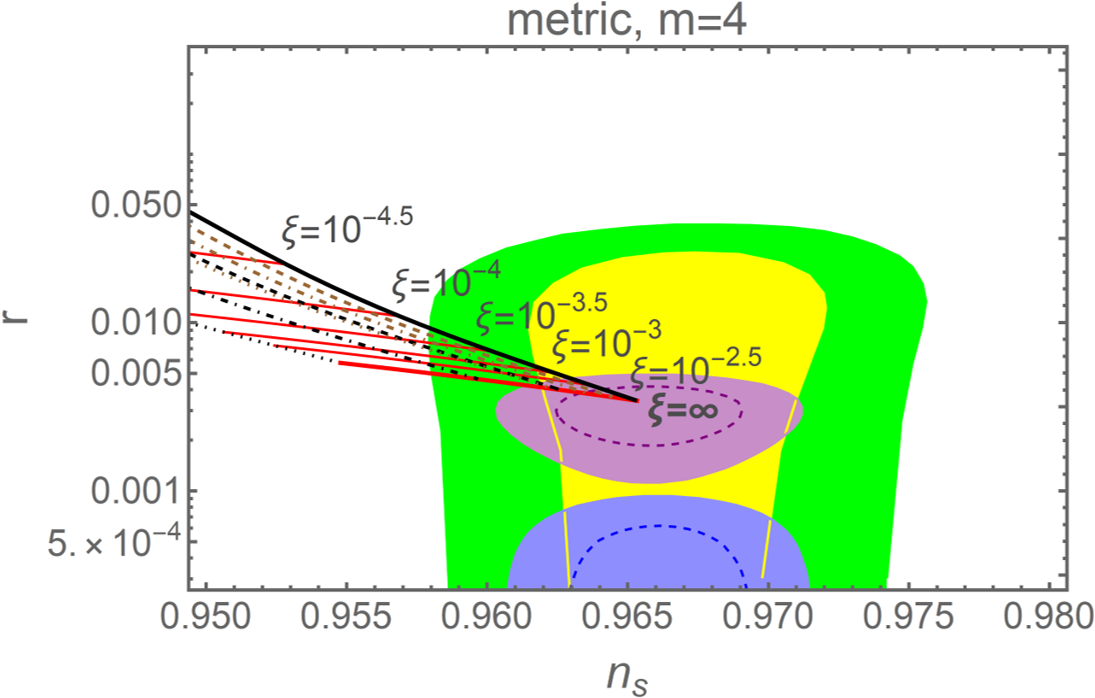}
	\includegraphics[width=15cm]{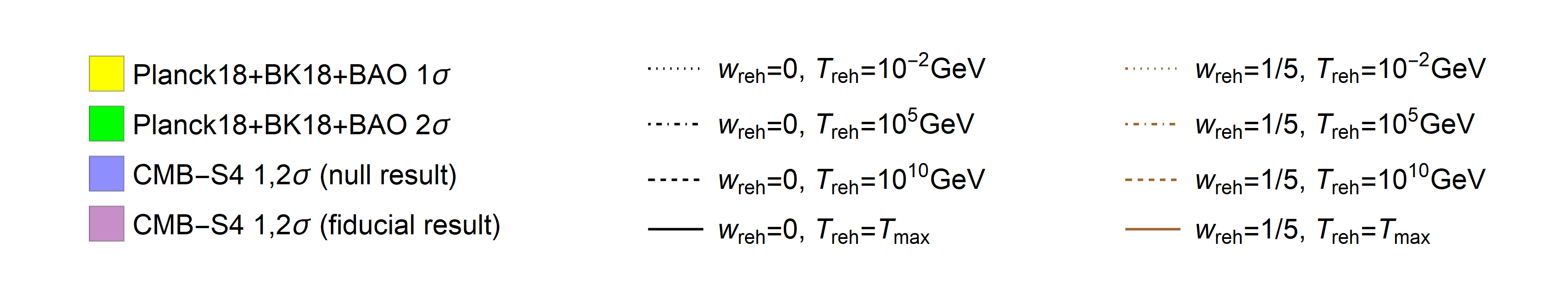}
	\caption{$ (n_{s},r) $ plot with $ m=(1,2,3,4) $ in metric formalism. Red solid lines shows the dependence of inflationary predictions on various $ \xi \geq 0$ values for $ w_{\text{reh}}\in \left[0, \frac{1}{3}\right] $ and $ T_{\text{reh}} \in \left[ T_{\text{BBN}},T_{\text{max}} \right] $. Dotted, dash-dotted, dashed, solid lines denote inflationary predictions with $ T_{\text{reh}} = (10^{-2},10^{5},10^{10},T_\text{max})~\text{GeV} $ respectively, along with $ w_{\text{reh}} = 0 $ for black and $ w_{\text{reh}} = \frac{1}{5} $ for brown lines.
	Current Planck18+BK18+BAO and expected future CMB-S4 measurements are denoted each in yellow/green and blue/purple. \label{Fig:metric}}
\end{figure}

\begin{figure}\centering
	\includegraphics[width=10cm]{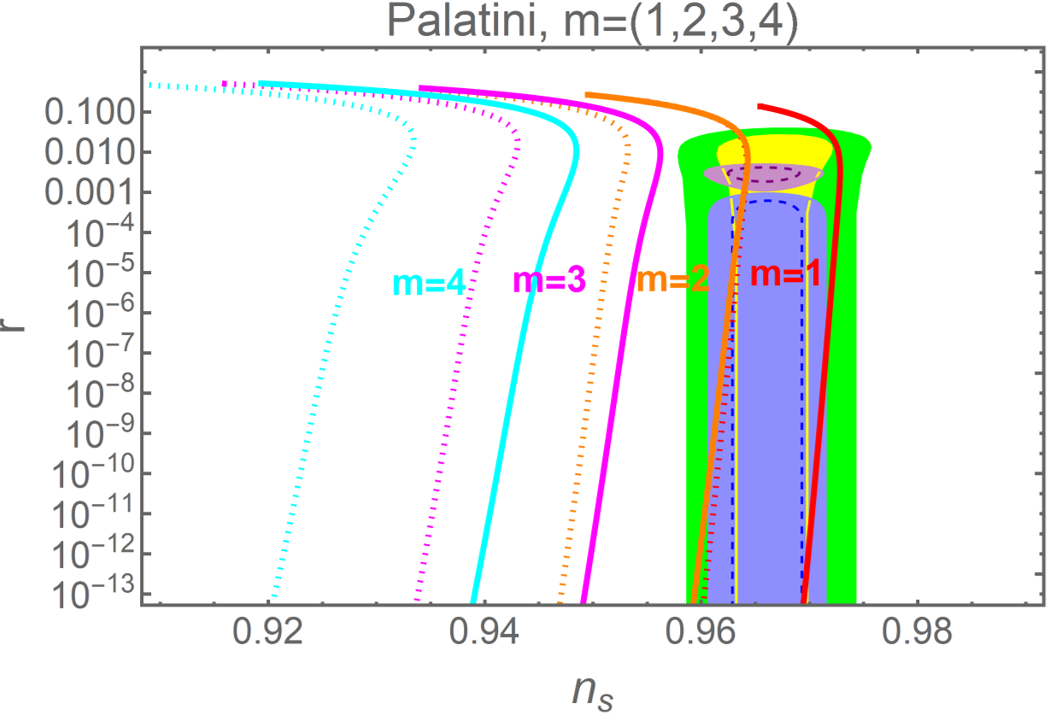}
	\caption{$ (n_{s},r) $ plot for $ m=(1,2,3,4) $ in Palatini formalism. Solid lines are for instantaneous reheating, and dotted lines are for $ T_{\text{reh}} = T_{\text{BBN}}  $. Except for $ m=1 $ and $ m=2 $, other power potentials are inconsistent to current observations, regardless of the reheating temperature for equation of state $ w_\text{reh}  \in \left[0 , \frac{1}{3}\right] $. \label{Fig:palatiniall}}
\end{figure}

\begin{figure}[t]
	\includegraphics[width=7.5cm]{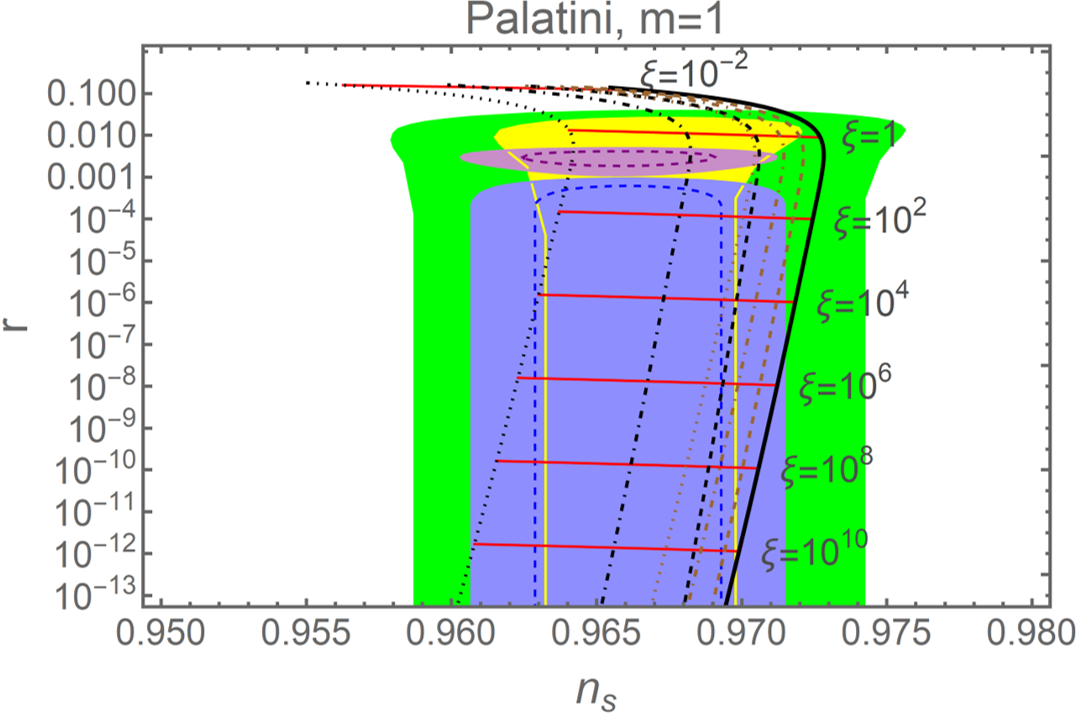}
	\includegraphics[width=7.5cm]{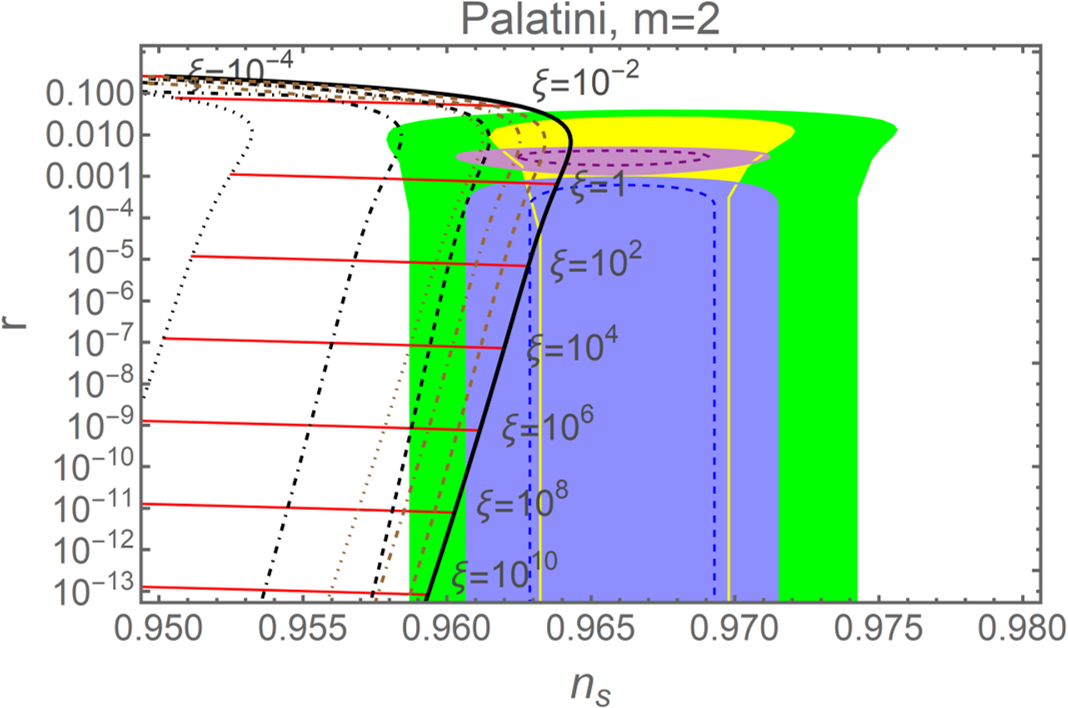}
	\includegraphics[width=15cm]{fig_metriclegends}
	\caption{$ (n_{s},r) $ plot for $ m=1 $ and $ m=2 $ in Palatini formalism.  Plotting conventions are the same as Figure.~\ref{Fig:metric}. Depending on the reheating temperature, predictions on $ n_{s} $ significantly changes as well. \label{Fig:palatini}}
\end{figure}

We present the exact results in Figure.~\ref{Fig:metric} for metric cases and Figure.~\ref{Fig:palatiniall}, Figure.~\ref{Fig:palatini} for Palatini cases. For all figures, we include Planck18+BICEP/Keck(BK)18+BAO $1 \sigma $ (yellow) and $2 \sigma$ (green) results  on $ (n_{s},r) $ as well as prospective results from future CMB-S4 observations with a fiducial detection of $ r = 3 \times 10^{-3} $ (purple) and null results (blue), both assuming a similar $ n_{s} $ center value with Planck18+BK18+BAO \cite{BICEPKeck:2021gln,Abazajian:2019eic}.

\begin{figure}
	\centering
	\includegraphics[width=15cm]{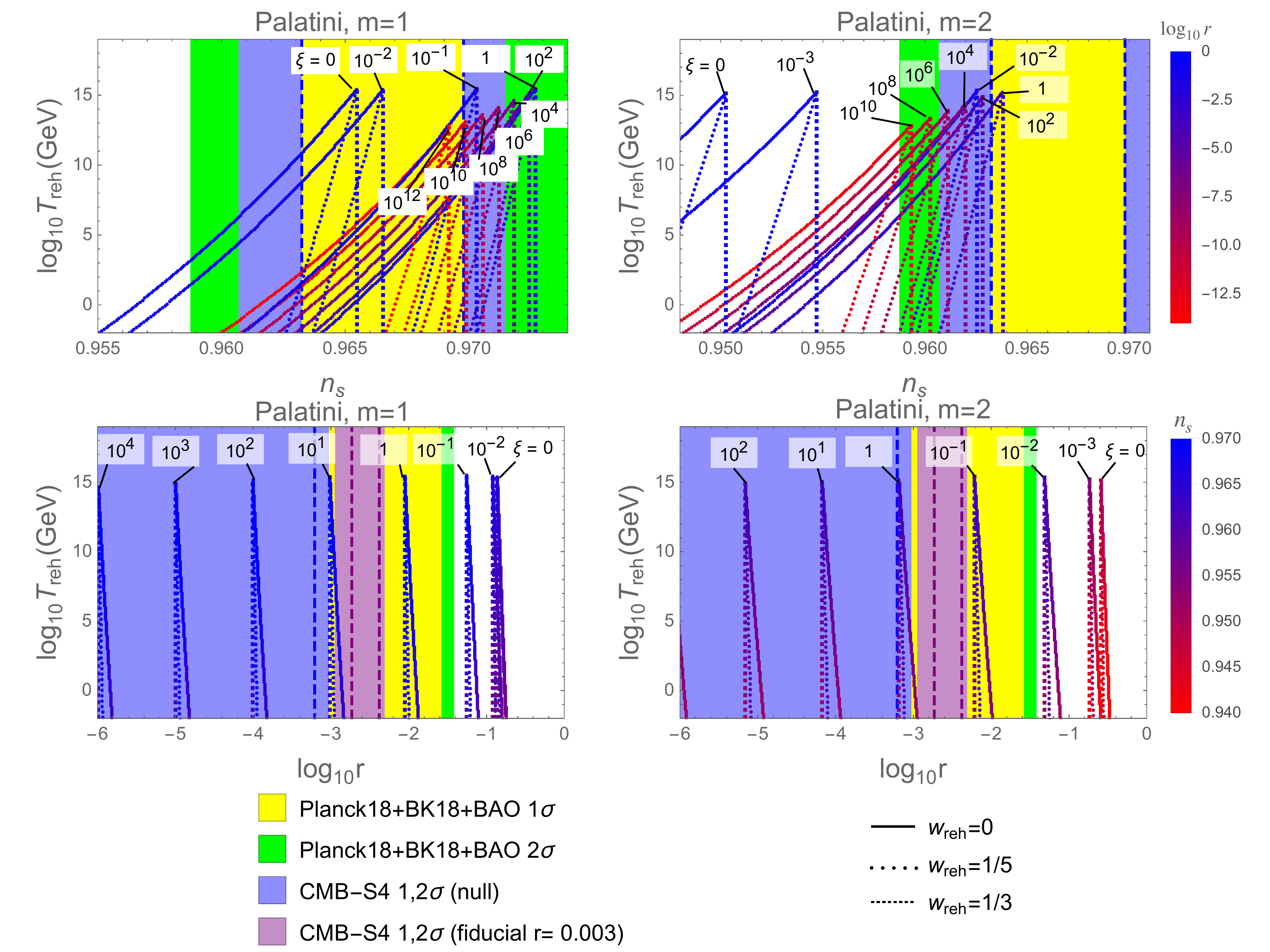}
	\caption{Constraints on the reheating temperature $T_\text{reh}$ depending on $n_s$ (top)  and $r$ (bottom) for the Palatini $m=1$ and $m=2$ cases. The spectrum of $r$ and $n_s$ is also depicted accordingly. The solid, point-dotted, square-dotted lines each correspond to $w_\text{reh}=0$, $w_\text{reh}= 1/5$, $w_\text{reh}= 1/3$, respectively. The Planck18+BK18+BAO $1 \sigma$ and $2 \sigma$ constraints are depicted in yellow and green, respectively. Future CMB-S4 null and fiducial constraints are also expressed in blue and purple, where the dashed lines and solid boundaries represent $1\sigma$ and $2\sigma$ bounds. For cases $m=3,4$, the predictions contradict with current CMB observations in terms of $n_s$, regardless of the reheating dynamics.
	\label{Figure:ConstraintsPala}}
\end{figure}

To begin with, we discuss how inflationary predictions differ in common in both metric and Palatini formalisms. First, the predictions of $ (n_{s}, r) $ change depending on $ (w_\text{reh},T_{\text{reh}}) $. For illustration, we denote predictions on $ (n_{s},r) $ for $ w_\text{reh}=0 $ (black) and $ w_{\text{reh}} = 1/5 $ (brown) depending on $ T_{\text{reh}} \in [10^{-2}~\text{GeV} - T_{\text{max}} ]$. $ T_{\text{max}} $ is determined by imposing instantaneous reheating
	\begin{align}
		\rho_{e} = \frac{3}{2}U_{e} = \frac{\pi^{2} }{30} g_\text{reh} T_{\text{max}}^{4}
	\end{align}
	at the end of the inflation. Lines corresponding to $ T_{\text{max}} $ are denoted by solid lines. For metric cases, $ T_{\text{max}} \simeq 10^{16}~\text{GeV} $ and for Palatini cases, $ T_{\text{max}} $ decreases for larger $ \xi $. (See Figure.~\ref{Figure:ConstraintsPala}.) On the other hand, the lower bound $ T_\text{reh} \sim 10^{-2}~\text{GeV} $ corresponds to the BBN scale temperature $ T_{\text{BBN}} $ as a conservative assumption and is depicted in dotted lines. Dot-dashed and dashed lines correspond to $ T_{\text{reh}} = 10^{5}~\text{GeV} $ and $ T_{\text{reh}} = 10^{10}~\text{GeV} $, respectively.\footnote{ 
		$ T_{\text{reh}} = 10^{5} ~\text{GeV} $ can be understood as an energy scale of future collider experiments for new physics searches. $ T_{\text{reh}} = 10^{10}~\text{GeV} $ is usually regarded as a typical upper bound of the reheating temperature of the `gravitino overproduction problem' if one takes this literally.} Lower reheating temperature gives smaller $ n_{s} $ predictions while the $ r $ dependence on reheating temperature is weak. Constraints on the reheating temperature from current/future $ n_{s} $ and $ r $ observations are also discussed below.
		 
Second, the regions of prediction become narrower and converge to the solid line on the right as $ w_\text{reh} \rightarrow 1/3 $. In this limit, the end of the reheating is ambiguous, because there is no distinct change in the equation of state compared to the radiation dominant phase. Instead, the predictions on $ (n_{s}, r) $ does not depend on reheating processes and is uniquely determined solely by inflationary model parameters.

We note that, depending on $ m $, even $ \xi \lesssim \mathcal{O}(1) $ still provides predictions compatible to current constraints. For example, in the metric formalism, intermediate $\xi$ values $\xi \sim \mathcal{O}(1)$ for $m=1$, and $\xi \sim \mathcal{O}(10^{-1})$ for $m = 2$ and each of these show different predictions for $ (n_{s},r) $, but still fully consistent to observations.

On the other hand, there are several notable differences between the two formalisms of gravity. For the Palatini case, each $ m $ give different predictions even in the $ \xi \rightarrow \infty $ limit. Potentials with the form $ m \geq 3 $ for $ w_\text{reh} \in \left[ 0 , \frac{1}{3} \right] $ are ruled out regardless of the reheating history. This is shown in Figure.~\ref{Fig:palatiniall}. (However, if one considers $w_\text{reh} > 1/3$, the  predictions shift to larger $n_s$ values so that these can be inside current observational bounds.)

Also, in the Palatini case, from the fact that $ r\vert_{\xi \rightarrow \infty} \propto 1/\xi $, the tensor-to-scalar ratio is highly suppressed, while for metric cases, predictions converge to specific values depending on $T_\text{reh}$ as long as $ \xi \gg 1 $. This was precisely why, Palatini Higgs inflation has hardly been regarded to be proven by observation in general. However, in this work, we show that the predictions of $ n_{s} $ for non-minimal inflation in the Palatini formalism actually possesses a  dependence on $ \xi $, when we take the reheating processes more carefully.

In the $m=1$ case, the predictions are still highly compatible with the CMB observations, regardless of the $w_\text{reh}$ value. Interestingly, for $ m=2 $ in the Palatini case, we observe that quartic potential inflation with a large non-minimal coupling $ \xi \gtrsim \mathcal{O}(10) $ is already outside of current Planck18+BK18+BAO $ 1 \sigma $ bounds, even for instantaneous reheating. (See the right of Figure.~\ref{Fig:palatini}.)  This also includes conventionally considered non-minimal coupling values for Palatini-Higgs inflation with $ \xi \gtrsim \mathcal{O}(10^{8}) $ (See Eq.~\eqref{Eq:PalaHiggsAmp} with  $ \lambda \gtrsim \mathcal{O}(10^{-2})  $). Also, $w_\text{reh}$ values deviating from $\frac{1}{3}$ tend to worsen the compatibility with observations for lower $T_\text{reh}$. For example, for $\xi = 10^8$,  current CMB bounds require $T_\text{reh} \gtrsim \mathcal{O}(10^{10})~\text{GeV} $ for $w_\text{reh} = 0 $ and  $T_\text{reh} \gtrsim  \mathcal{O} (10^5)~\text{GeV} $ for $w_\text{reh}= \frac{1}{5} $. These constraints will be more stringent for future observations such as CMB-S4 \cite{Abazajian:2019eic} and LiteBIRD \cite{LiteBIRD:2020khw}, lying beyond the $2\sigma$ expected bounds.
This fact also can be used to distinguish metric Higgs inflation models from Palatini cases \cite{Bauer:2010jg, Rasanen:2017ivk, Markkanen:2017tun, Jinno:2019und, Shaposhnikov:2020fdv, Gialamas:2020vto, Enckell:2020lvn}. Note that the early stage of the reheating (namely `preheating') of Palatini Higgs inflation possesses $ w \simeq -1 $ \cite{Rubio:2019ypq}, which makes the compatibility to the observations worse if the reheating processes has an averaged equation of state $ w_\text{reh} < 1/3 $.\footnote{As the reheating stage of the Palatini Higgs inflation is tachyonic, we expect the reheating to be very efficient so that $ w_\text{reh} $ is very close to $ 1/3 $. However, detailed analysis on the reheating of this model and determination of the precise $ w_\text{reh} $ is out of the scope of this work.}

\begin{figure}[t]
	\centering
	\includegraphics[width=15cm]{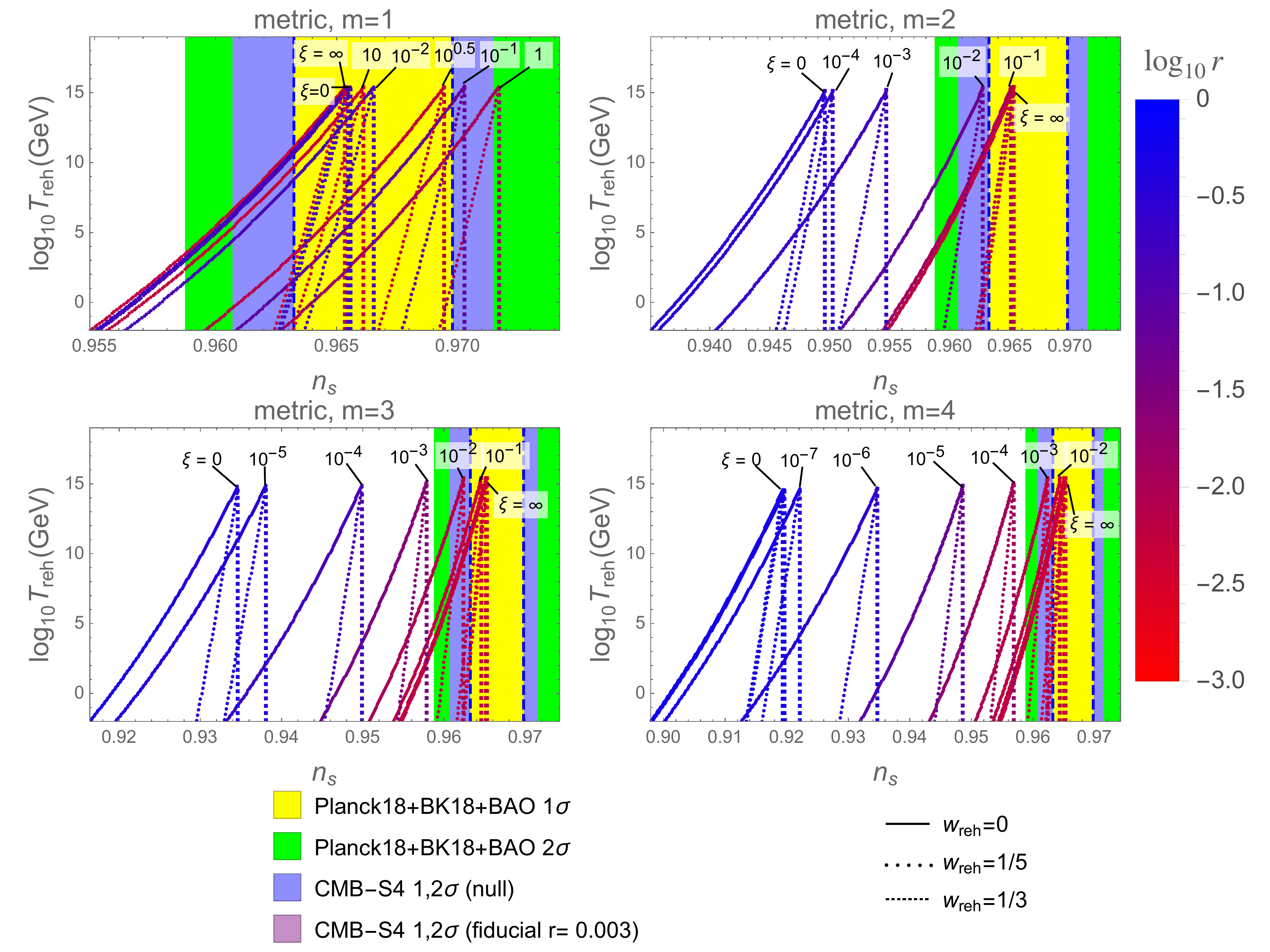}
	\caption{Constraints on the reheating temperature $T_\text{reh}$ depending on $n_s$ for the metric $m=(1,2,3,4)$ cases. The spectrum of $r$ is also depicted. The solid, point-dotted, square-dotted lines each correspond to $w_\text{reh}=0$, $w_\text{reh}= 1/5$, $w_\text{reh}= 1/3$, respectively. The Planck18+BK18+BAO $1 \sigma$ and $2 \sigma$ constraints are depicted in yellow and green, respectively. Future CMB-S4 null and fiducial constraints are also expressed in blue and purple, where the dashed lines and solid boundaries represent $1\sigma$ and $2\sigma$ bounds. %Note that these observables can provide a lower bound on $T_\text{reh}$. 
		\label{Figure:Constraints_n_s}}
\end{figure}

\begin{figure}
	\centering
	\includegraphics[width=15cm]{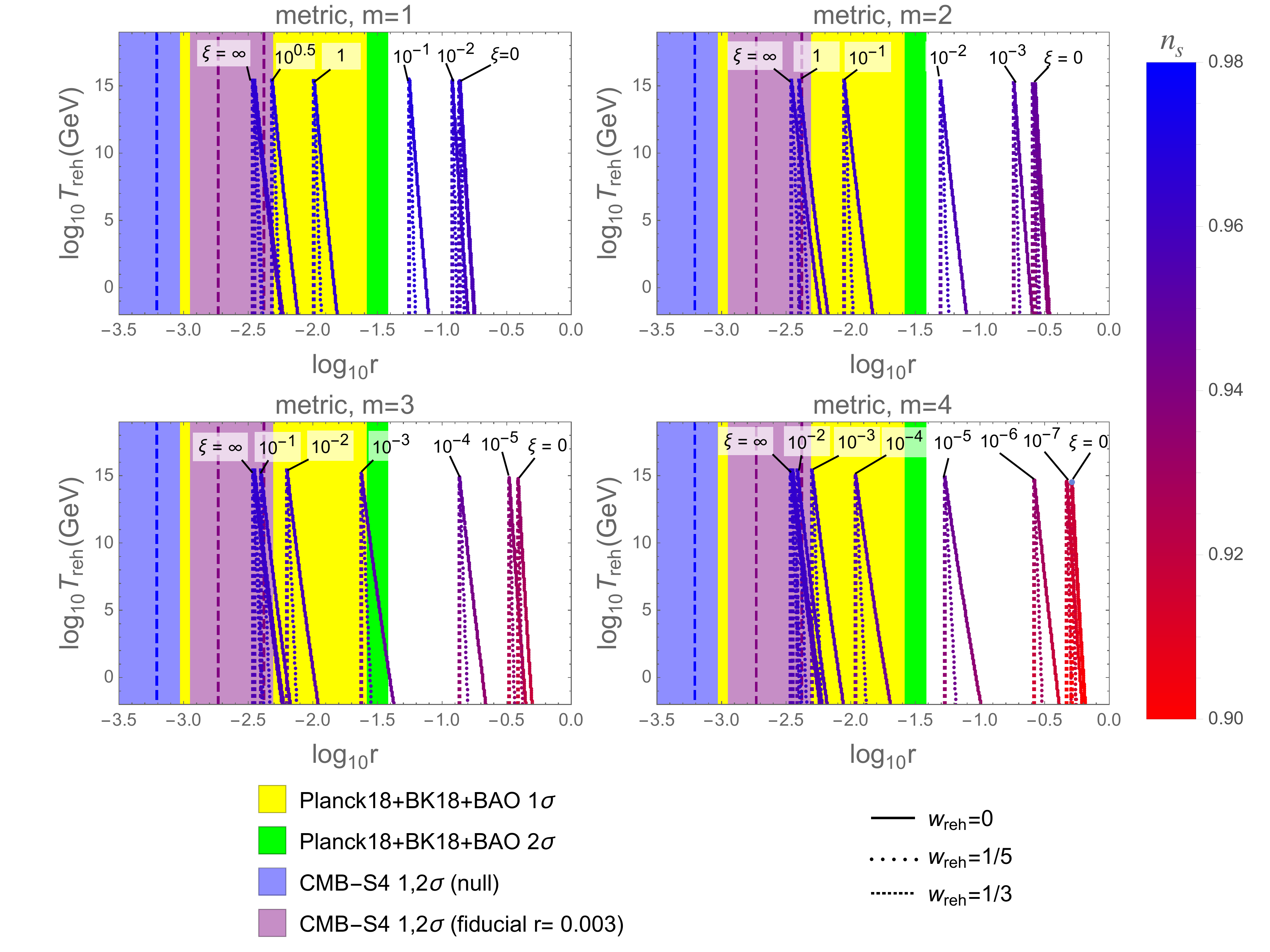}
	\caption{Constraints on the reheating temperature $T_\text{reh}$ depending on $r$ for the metric $m=(1,2,3,4)$ cases. The spectrum of $n_s$ is also depicted. The solid, point-dotted, square-dotted lines each correspond to $w_\text{reh}=0$, $w_\text{reh}= 1/5$, $w_\text{reh}= 1/3$, respectively. The Planck18+BK18+BAO $1 \sigma$ and $2 \sigma$ constraints are depicted in yellow and green, respectively. Future CMB-S4 null and fiducial constraints are also expressed in blue and purple, where the dashed lines and solid boundaries represent $1\sigma$ and $2\sigma$ bounds.
	\label{Figure:Constraints_r}}
\end{figure}

%\subsection{Constraints on Reheating Temperature}
%\label{Section:Constraints on Reheating Temperature}
Following the preceding results, bounds on $ (n_{s}, r) $ also impose constraints on $T_\text{reh}$ depending on $ w_{\text{reh}} $. Relations between $(n_s, T_\text{reh}) $ for metric cases are depicted in Figure.~\ref{Figure:Constraints_n_s}. The $\xi$ values range from $\xi =0 $ to $\xi = \infty $ for all $m=(1,2,3,4)$ potentials. The allowed range of $T_\text{reh}$ varies depending on the potential form along with the specific non-minimal coupling value. We also note that, for $ m=1 $, there exists some range of $ \xi $ which gives upper bound on $ T_{\text{reh}} $ as well depending on $ w_{\text{reh}} $. The $(r, T_\text{reh}) $ relation for metric cases is also shown in Figure.~\ref{Figure:Constraints_r}. As the prospected sensitivity at CMB-S4 improves significantly compared to Planck18+BK18+BAO up to $r\lesssim10^{-3}$, these observations may be used to effectively constrain $T_\text{reh}$.

The $T_\text{reh}$ implications for the Palatini case with $ m = (1,2) $, which is depicted in Figure.~\ref{Figure:ConstraintsPala}, differ from the metric scenario. The suppression of $r \propto 1/\xi$ in the Palatini formalism makes it nearly impossible for future observations to provide reasonable constraints on the allowed $T_\text{reh}$ range for large $\xi$ parameters. Instead, the $\xi$ dependence on $n_s$ may lead to providing a more stringent (upper/lower) bound on $T_\text{reh}$.

\section{Conclusion and Discussions}
\label{Section:Conclusion and Discussions}

In this work, we clarified the effects of reheating to predictions of inflation models with non-minimal coupling to gravity, covering (i) both metric and Palatini formalisms, (ii) arbitrary monomial potential with asymptotic flatness in the Einstein frame, and (iii) a wide range of non-minimal coupling $ \xi $ depending on the power of the potential, from the consistency relations between reheating parameters and inflationary observables $ n_{s} $ and $ r $. We also obtained a range of compatible reheating temperatures for a given equation of state parameter $w_\text{reh}$ during reheating. This in turn can be used complementarily to further decipher the microscopic particle physics governing the reheating process for each model.

For models in both formalisms, we note that the CMB compatible $\xi$ values also allow $\xi \leq 1$, with its predictions deviating from those for large $\xi \gg 1$ limits. The specific dependence on $(n_s, r)$ in this particular $\xi$ range differ with a definite dependence on the details of the model's potential.

Especially, for $m=2$ in the Palatini formalism, while $\xi \gtrsim \mathcal{O}(10^{-2})$ are allowed for current CMB $2\sigma$ bounds, the dependence on $\xi$ indicates that typically considered large $\xi \sim \mathcal{O}(10^8)$ lead to $N_k$ smaller than 50, which is not preferred within the $1\sigma$ limit of current CMB bounds. Current and future CMB observations are able to constrain the particular parameter range of $\xi$s, giving definite lower and upper bounds on the non-minimal coupling, with the constraints being stronger when taking the effects of reheating into account.

We emphasize that our results are applicable to a wider class of models, which exhibit $\alpha$-attractor behavior. $ R^{2} $-driven Starobinsky inflation is also transformable to this class, giving the same inflationary dynamics.  Application to multi-field inflationary models that are reducible into an effective single-field $\alpha$-attractor case is also straightforward. However, the reheating details of these equivalent inflationary classes differ depending specifically on the model that describes its microscopic nature, and predicts different values of $(T_{\text{reh}}, N_{\text{reh}}, w_{\text{reh}}) $. Therefore, our result could be used as a generic template for the inflationary predictions of wide classes of models, breaking the degeneracy of the various models not caught solely from considering the inflation dynamics. This approach will provide valuable information on deciphering the contents of the post-inflationary universe, which may feature - but most definitely not restricted to - multiple fields encompassing spectator fields, new interactions including higher order gravitational couplings.\footnote{Single field inflation models including higher order $R^{3}$  have been analyzed in Ref. \cite{Cheong:2020rao}.} We will further investigate this direction in future works.

\acknowledgments
We thank Fedor Bezrukov, Kin-ya Oda, and Minxi He for discussions throughout various stages of the research. This work is supported in part by the National Research Foundation of Korea (NRF) grant funded by the Korea government (NRF-2019R1A2C1089334), (NRF-2021R1A4A2001897) (SCP), and (NRF-2020R1A6A3A13076216) (SML). SML is supported by the Hyundai Motor Chung Mong-Koo Foundation Scholarship.

\appendix


\bibliographystyle{utphys}
\bibliography{nonminimalreheating.bib}

\end{document}